\PassOptionsToPackage{table}{xcolor}
\documentclass[acmtog,nonacm]{acmart}
\setcitestyle{square}
\setcopyright{none}
\acmJournal{TOG}
\acmYear{2022}\acmVolume{41}\acmNumber{6}\acmArticle{227}\acmMonth{12} \acmDOI{10.1145/3550454.3555441}

\author{Silvia Sell\'{a}n}
\email{sgsellan@cs.toronto.edu}
\affiliation{%
  \institution{University of Toronto}
}
\author{Alec Jacobson}
\email{jacobson@cs.toronto.edu}
\affiliation{%
  \institution{University of Toronto}
}
\affiliation{\institution{Adobe Research}
}

\usepackage[mathletters]{ucs}
\usepackage[utf8x]{inputenc}

\newcommand{\refequ}[1]{Eq.~(\ref{equ:#1})}
\newcommand{\refapp}[1]{Appendix \ref{app:#1}}
\newcommand{\reffig}[1]{Fig.~\ref{fig:#1}}
\newcommand{\refsec}[1]{Section~\ref{sec:#1}}

\providecommand{\C}{}
\providecommand{\D}{}
\providecommand{\E}{}

\providecommand{\K}{}
\providecommand{\L}{}

\providecommand{\N}{}

\providecommand{\R}{}
\providecommand{\S}{}

\providecommand{\V}{}

\providecommand{\Z}{}
\providecommand{\a}{}

\providecommand{\k}{}

\providecommand{\m}{}

\providecommand{\x}{}

\renewcommand{\R}{\mathbb{R}}
\renewcommand{\x}{x}

\renewcommand{\N}{\mathcal{N}}
\renewcommand{\S}{\mathcal{S}}
\renewcommand{\L}{\mathbf{L}}
\renewcommand{\V}{\mathbf{V}}
\renewcommand{\D}{\mathbf{D}}
\renewcommand{\C}{\mathbf{C}}
\renewcommand{\k}{\mathbf{k}}
\renewcommand{\K}{\mathbf{K}}
\renewcommand{\Z}{\mathbf{Z}}
\renewcommand{\E}{\mathbf{E}}

\renewcommand{\m}{\mathbf{m}}
\renewcommand{\a}{\mathbf{a}}

\DeclareMathOperator{\diag}{diag}
\definecolor{derekTableBlue}{rgb}{0.565,0.847,0.769} 
\DeclareRobustCommand{\rchi}{f}

\usepackage{tabularx}
\usepackage{booktabs}
\usepackage{makecell}
\usepackage[table]{xcolor}
\definecolor{white}{rgb}{1,1,1}
\definecolor{lightbluishgrey}{rgb}{0.76471,0.84824,0.91647}
\definecolor{newcolor}{HTML}{006633}

\usepackage{subcaption}
\usepackage{wrapfig}
\usepackage{graphicx}

\usepackage{listings}
\usepackage{layouts}
\makeatletter 
\newcommand{\layoutdetails}{%
\begin{tabular}{ll}
 \texttt{\textbackslash{textwidth}} & \printinunitsof{in}\prntlen{\textwidth} \\
\texttt{\textbackslash{linewidth}} & \printinunitsof{in}\prntlen{\linewidth} \\
Main text font &  \f@size pt \f@family \\
\sffamily \small Caption text font &  \sffamily \small \f@size pt \f@family \\
\end{tabular}%
}
\makeatother

\newcommand{\new}[1]{{#1}\normalfont}

\begin{document}

\title{Stochastic Poisson Surface Reconstruction}

\begin{abstract}
    We introduce a statistical extension of the classic Poisson Surface
    Reconstruction algorithm for recovering shapes from 3D point clouds. Instead
    of outputting an implicit function, we represent the reconstructed shape as a
    modified Gaussian Process, which allows us to conduct statistical queries
    (e.g., the likelihood of a point in space being on the surface or inside a
    solid). We show that this perspective: improves PSR's integration into the
    online scanning process, broadens its application realm, and opens the door to other
    lines of research such as applying task-specific priors.
\end{abstract}

\begin{teaserfigure}
    \includegraphics{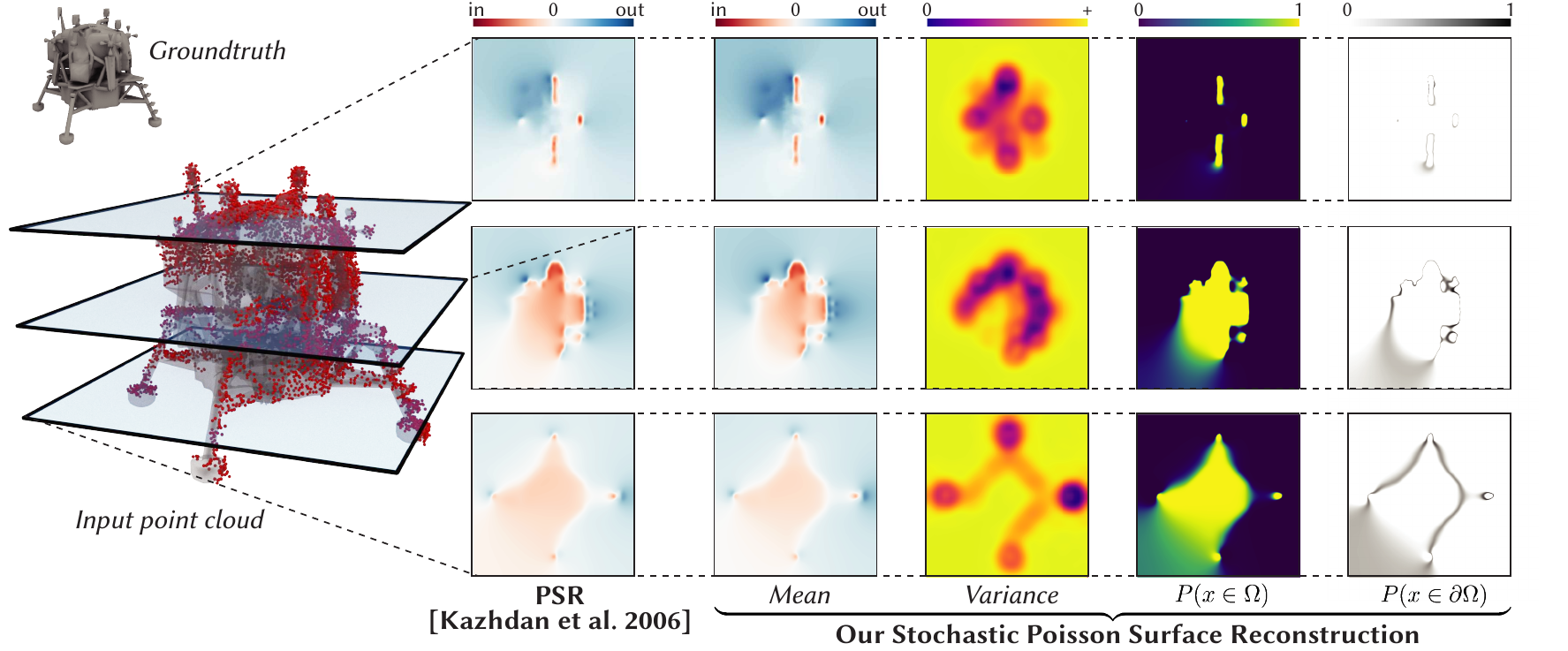}
    \caption{\new{Left to right: given an input point cloud, Poisson Surface Reconstruction (PSR) recovers the surface as the zero levelset of an implicit function. We propose a novel statistical derivation of PSR that exchanges the function for a distribution, allowing us to answer many statistical queries.}}\label{fig:teaser}
\end{teaserfigure}

\maketitle

\section{Introduction}

Surface reconstruction refers to the process of converting a  point cloud (the
most common real-world raw 3D capture format) into another shape representation,
such as a mesh or an implicit function, for use in downstream applications. 
This is an \emph{underdetermined} process filled with \emph{uncertainty}, not
just due to a point cloud's discrete nature and lack of topological information
but also because of real-world challenges like scan occlusions or measurement
error.

The \emph{de facto} standard geometry processing algorithm for this task is
\emph{Poisson Surface Reconstruction} (PSR) \cite{kazhdan2006poisson}, which
solves a partial differential equation to reconstruct a function $\rchi_{PSR}$
whose zero levelset $\rchi_{PSR}=0$ defines the desired surface. Due to its
speed, quality and simplicity, PSR remains relevant and has seen uses in fields
as varied as digital heritage preservation \cite{andrade20123d}, topography
\cite{gupta20173d}, medicine \cite{palomar2016surface} and autonomous driving
\cite{vizzo2021poisson}.

\begin{figure}[b]
    \centering
    \includegraphics{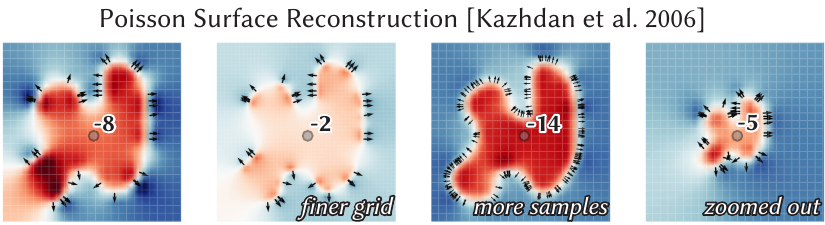}
    \caption{PSR values outside of zero are arbitrary (e.g., they depend on grid size, sampling rate and scale) and contain no direct statistical information.}\label{fig:psr-bad}
\end{figure}

Unfortunately, PSR lacks the statistical formalism to quantify the uncertainties
of the surface reconstruction process. The magnitudes of $\rchi_{PSR}$ outside
of the zero levelset are arbitrary (see \reffig{psr-bad}) and $\rchi_{PSR}$
alone cannot provide an answer to statistical questions crucial to the
reconstruction process like ``how confident can one be of the values of
$\rchi_{PSR}$?'' or ``where should one aim the scanner next to optimize
information gain?'' Similarly, it cannot respond to queries like ``what is the
probability of a point $p$ being contained in the shape?'', critical for
collision detection or ray casting applications.

In this paper, we introduce \emph{Stochastic Poisson Surface Reconstruction} (SPSR), a statistical derivation of PSR as conditional probability distributions in a Gaussian Process (GP). Instead of just an implicit function value, we endow every point in space with a Gaussian distribution of possible values. 
We propose an algorithm to compute the mean and variance that fully determine this distribution (see \reffig{teaser}), allowing us to answer any statistical queries.

This extension vastly broadens the use cases of Poisson Surface Reconstruction, as we highlight with prototypical examples of surface point cloud repair, ray casting, next-view planning and collision detection. Furthermore, we show that by understanding PSR \new{from this new perspective}, we can borrow from the Gaussian Process literature to modify it by incorporating task-specific priors, opening several promising lines of future research.

\section{Related Work}

\begin{figure}
    \centering
    \includegraphics{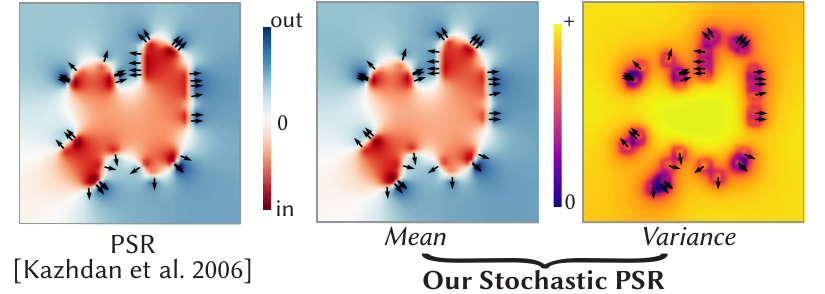}
    \caption{\new{Our Stochastic PSR extends the traditional PSR into a statistical distribution whose mean is nearly identical to the PSR output.}}\label{fig:psr-vs-spsr}
\end{figure}

A complete survey of vast research areas like uncertainty quantification or surface reconstruction is beyond the scope of this work. Instead, we focus this section on setting our Stochastic PSR in its context of bridging the gap between PSR and Gaussian Processes.

\subsection{Surface Reconstruction from Point Clouds}\label{sec:surfacereconstruction}

Point clouds are a common raw format for 3D geometry acquired form the real world. However, most applications in fields like rendering, simulation and geometry processing require more structured representations like triangle meshes. Thus, reconstructing surfaces from point clouds is a well-studied, fundamental problem in Computer Graphics (see \cite{berger2017survey} for an exhaustive survey). While some methods convert point clouds directly to meshes (e.g., by dictionary learning \cite{xiong2014robust}) or simple primitives (e.g., \cite{monszpart2015rapter,nan2010smartboxes}), we focus on those that extract the shape as the zero levelset of a reconstructed function $\rchi$.

Within these, a common separation is made between \emph{local} and \emph{global}
algorithms. Local algorithms prioritize performance in speed and memory; for
example, by fitting linear \cite{hoppe1992surface}, polynomial
\cite{alexa2001point} or higher-order \cite{fuhrmann2014floating} functions to
restricted point subsets. By their nature, these are more susceptible to
oscillations far from the sample points.  To cope, global algorithms (e.g.,
\cite{jacobson13winding}) allow $\rchi$ to be influenced by every point in the
cloud (hierarchical fast summation structures can help reduce computation
\cite{barill2018fast}).

Poisson Surface Reconstruction (PSR) \cite{kazhdan2006poisson} 
captures the best features of the global (robustness) and local (performance)
methods by computing $\rchi$ in two steps. First, a vector field $\vec{V}$ is
interpolated from the point cloud using only local information. Then, $\rchi$ is
obtained from $\vec{V}$ via a global sparse PDE solve, which is discretized as
a linear system and can be solved very efficiently using an adaptative grid
structure. PSR has been improved since its publication; for example,
\citet{kazhdan2013screened} combine both steps to improve noise robustness,
\citet{kazhdan2020poisson} include envelope constraints and
\citet{peng2021shape} formulate the Poisson solve in a differentiable way.
Nonetheless, the original 2006 publication is still one of the few showing
robustness in every metric considered by \citet{berger2017survey} more than a decade 
later.

Our Stochastic PSR inherits all the benefits of the original PSR, supplying it
with a complete statistical formalism that extends it and its application realm
(see \reffig{psr-vs-spsr}). Our contributions are orthogonal to the specific
grid structure used (see \cite{kazhdan2019adaptive}).

\begin{figure}
    \centering
    \includegraphics{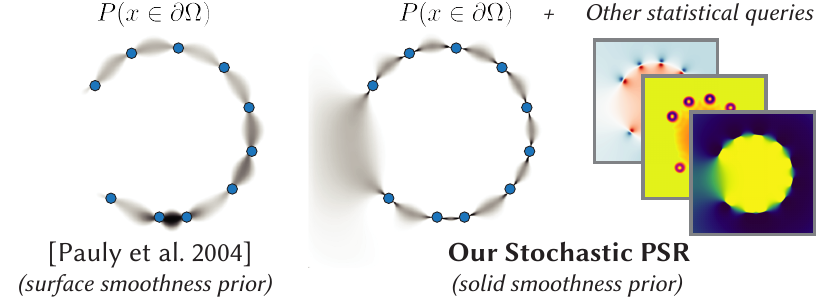}
    \vspace{-0.3cm}
    \caption{\citet{pauly2004uncertainty} use a surface smoothness prior to quantify uncertainty in reconstruction. We use a \emph{solid} smoothness prior to compute a full statistical distribution, from which we can also query surface quantities.}\label{fig:pauly}
\end{figure}

Our algorithm can output the probability of any point in space being inside the
sampled domain (see \reffig{teaser}). While this resembles the shape
representations proposed by \emph{occupancy networks}
\cite{mescheder2019occupancy} and \emph{neural radiance fields}
\cite{mildenhall2020nerf}, we note that our algorithm produces this
quantity as a direct byproduct of a fully determined statistical distribution which can
answer many other statistical queries, like boundary probabilities (see
\reffig{pauly}) or regional probabilities (see \reffig{integrate-region-full}).

\begin{figure}
    \centering
    \includegraphics{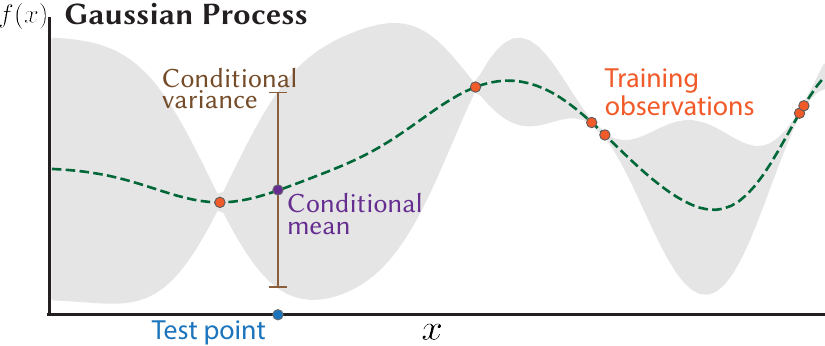}
    \caption{A sample Gaussian Process applied to a supervised learning task. Given some training observations (orange), any unobserved test point is given a conditional distribution with a mean (purple) and variance (brown).}\label{fig:implicit-gaussian-didactic}
\end{figure}

\vspace{-0.1cm}
\subsection{Gaussian Processes}

A Gaussian Process (GP) is an infinite collection of joint normal distributions
\cite{doob1944elementary,dudley2018real}, usually parametrized by a continuous
parameter like time or space \cite{kac1947theory}. Recently, Gaussian processes
have been used as a tool in unsupervised learning (see
\cite{williams2006gaussianbook} for an introduction,
\cite{engel2005reinforcement,raissi2017machine} for examples), even suggested
initially as an alternative to neural networks by \citet{mackay1997gaussian}.

Closer to our application are Gaussian Process Implicit Surfaces (GPIS) \cite{williams2006gaussian}. These algorithms exchange the $\rchi(\x)$ function that implicitly defines a surface for a Gaussian distribution $\rchi(\x)\sim\mathcal{N}(\mu(x),\sigma(x))$, and compute $\mu(x)$ and $\sigma(x)$ by studying the posterior GP distribution given observed points. One can recover a surface by extracting the zero levelset of $\mu$; however, the information contained in $\sigma$ also finds use in tasks like robotic grasping \cite{dragiev2011gaussian}, next view planning \cite{hollinger2012uncertainty} and segmentation \cite{shin2017real,ramon2017multi}.

GPIS present the same global/local dilemma as other surface reconstruction algorithms. If the interpolation is done using all the point cloud information (this is encoded in the support of the GP \emph{covariance} function), the method suffers in performance; if not, in robustness. By using the insights of Poisson Surface Reconstruction, our Stochastic PSR instead uses a local Gaussian Process to build the distribution of the gradient field $\nabla \rchi (x) $, and then recover the full distribution of $\rchi(x)$ with a global PDE solve.
While using a GP to interpolate vector-valued data is unorthodox, we note it has been done before; e.g., for fluid velocity information \cite{lee2019online}.

\new{
Gaussian processes are a more traditional approach to quantifying the uncertainty of a regression model, a field which has seen significant growth since recent advances in deep learning (see \cite{abdar2021review} for a survey). Often, the posterior distribution of a given neural network is approximated by another \emph{Bayesian} network whose parameters minimize a chosen distribution loss. (see e.g., \cite{xue2019reliable}). Alternatively, the posterior may be learned directly from the observed data (see e.g., \cite{shen2021stochastic}).
} 

\begin{figure}
    \centering
    \includegraphics{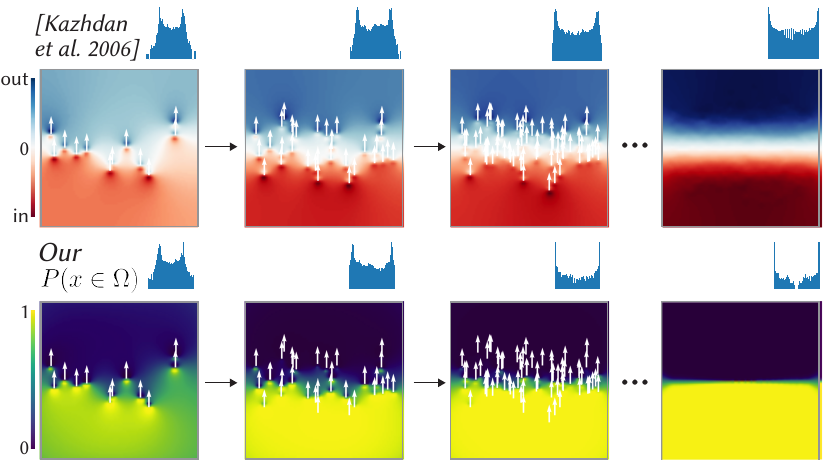}
    \caption{\new{Our SPSR provides a probability that accounts for sampling and cannot be
    recovered from the PSR values. Histograms in logarithmic scale.}}\label{fig:convergence}
\end{figure}

\vspace{-0.1cm}
\subsection{Stochastic Geometry Processing}

Our algorithm stands among many similar works that add statistical formalism to standard geometry processing techniques, like point cloud registration. 
Specifically, a lot of work has been dedicated to computing the covariances in the pairwise iterative closest point \cite{landry2019cello,bosse2008map} and, most recently, for general multi-scan registration \new{\cite{cao2018real,huang2020uncertainty}}. 

Closer to our application, \new{\citet{curless1996volumetric} compute truncated signed distance fields for each sensor position and use a model for their individual uncertainty to weigh their contributions to the final implicit reconstruction.}
\citet{pothkow2011probabilistic} model grid samples of an implicit function as normal distributions and calculate the individual probabilities of each voxel marching cubes configuration to quantify the contouring uncertainty. \new{\citet{sharf2007interactive} similarly measure topological reconstruction uncertainty to identify where user-provided disambiguation is most needed.}

\citet{pauly2004uncertainty} work is the most similar to ours. It takes a point
cloud as input and uses a surface smoothness prior to compute the likelihood of
any point in space lying on it (see \reffig{pauly}, left).  Instead, our
proposed algorithm assumes samples to lay on the boundary of a solid, and impose
solid smoothness prior (examined further in \refsec{beyondspsr}). Further, while
our algorithm can also output a surface likelihood quantity (see \reffig{pauly},
center), it is only part of a full statistical distribution of the solid (see
\reffig{pauly}, right).

\section{Background}\label{sec:background}

By posing PSR as a Gaussian Process, our work
combines these two concepts. We begin by reviewing them individually.

\begin{figure}
    \centering
    \includegraphics{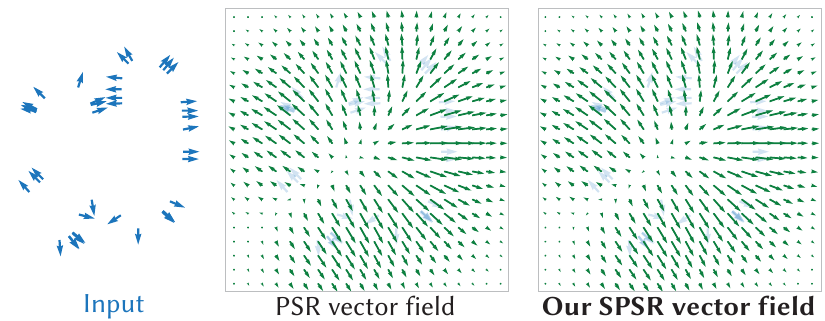}
    \caption{Our SPSR vector field $\vec{V}_{SPSR}$ (right), which uses a
    symmetrized version $k_{SPSR}$ of the traditional PSR covariance $k_{PSR}$, is
    visually identical to the PSR vector field $\vec{V}_{PSR}$ (center) for a
    representative input point cloud (left). \new{In the language of \refsec{spsr}, the right-most subfigure shows the mean of our vector field Gaussian Process $\vec{V}(q)$ after covariance lumping.}}\label{fig:vector-field-didactic}
\end{figure}

\subsection{Gaussian Processes}\label{sec:gp}

Intuitively, a Gaussian Process is an extension of the multivariate normal
distribution to an infinite number of dimensions. Formally, let
$\mathcal{A}=\{A(\x)\}_{\x\in D}$ be a collection of random variables
parametrized by some continuous parameter $\x$. $\mathcal{A}$ is said to be a
Gaussian Process if any finite subset of $\mathcal{A}$ follows a multivariate
Gaussian distribution. Equivalently, $\mathcal{A}$ is a Gaussian Process if for
any two $\x,\x'\in\Omega$,
\begin{equation}
    A(\x),A(\x') \sim \N\left( \begin{bmatrix}
        m(\x)        \\
        m(\x')
    \end{bmatrix},  \begin{bmatrix}
        k(\x,\x)   &  k(\x,\x')    \\
        k(\x',\x) & k(\x',\x')
    \end{bmatrix} \right)
\end{equation}
for some \emph{mean} and \emph{covariance} functions $m:\Omega\rightarrow\R$,
$k:\Omega\times \Omega\rightarrow \R$. These two functions uniquely determine
the Gaussian Process $\mathcal{A}$.

Gaussian Processes are a particularly useful tool for supervised learning tasks
(see \reffig{implicit-gaussian-didactic}). Assume training observations
$\mathcal{T}=\{(\x_i,a_i)\}_{i=1}^n$, where each $a_i$ is an observation of the
distribution $A(\x_i)$, and consider an unobserved (test) point $\x$. By
definition of GP, the joint distribution of $\{A(\x),A(\x_1),\dots,A(\x_n)\}$ is
{\normalsize
\begin{equation}
    \N\left( \begin{bmatrix}
        m(\x)        \\
        m(\x_1)        \\
        \vdots \\
        m(\x_n)
    \end{bmatrix},  \begin{bmatrix}
        k(\x,\x )   &  k(\x ,\x_1) & \dots & k(x ,\x_n)    \\
        k(\x_1,\x )   &  k(\x_1,\x_1) & \dots & k(\x_1,\x_n)  \\
        \vdots   &  \vdots & \ddots & \vdots  \\
        k(\x_n,\x )   &  k(\x_n,\x_1) & \dots & k(\x_n,\x_n)  
    \end{bmatrix} \right)
\end{equation}}
which we write as 
\begin{equation}
    A(\x),A(\x_1),\dots,A(\x_n) \sim \N\left( \begin{bmatrix}
        m_1        \\
        \m_2     
    \end{bmatrix},  \begin{bmatrix}
        k_1   &  \k_2^\top    \\
        \k_2    &  \K_3  
    \end{bmatrix} \right)\, ,
\end{equation}
where it is relevant to note that $\K_3$ depends only on the training data and
$\k_2$ depends on both training and test sets. By Bayes' theorem, this means the
distribution of $A(x)$ conditioned on the observations $\{(\x_i,a_i)\}_{i=1}^n$
is
\begin{equation}
    A(\x)\,|\,\mathcal{T} \sim \N (m_1 + \k_2^\top \K_3^{-1} (\a - \m_2),k_1 - \k_2^\top \K_3^{-1} \k_2)
\end{equation}
One usually assumes $m=0$ (otherwise, the Gaussian process $\mathcal{A}-m$ is
considered), and writes
\begin{equation}\label{equ:conditionalprobability}
    A(\x)\,|\,\mathcal{T}\sim \N (\k_2^\top \K_3^{-1} \a,k_1 - \k_2^\top \K_3^{-1} \k_2)\, .
\end{equation}
In the case of noisy observations, a term $\sigma_n^2 \mathbf{I}$ (where
$\sigma^2_n$ is the \emph{noise variance}) is added to $\K_3$.

\begin{figure}
    \centering
    \includegraphics{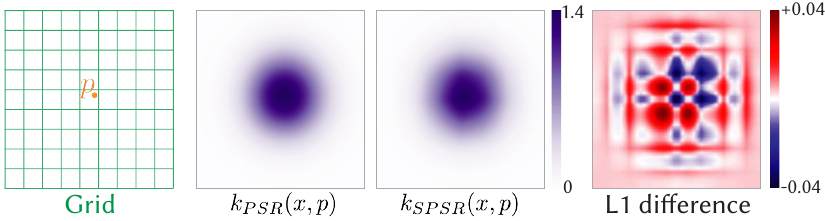}
    \caption{To interpret PSR as a Gaussian Process, we define $k_{SPSR}$, a minor
    modification of the PSR semicovariance $k_{PSR}$. These are visually similar
    and their difference is small (a maximum of
    $2\%$). }\label{fig:covariance-didactic}
\end{figure}

\subsection{Poisson Surface Reconstruction}

Any input oriented point cloud of the surface of a solid shape $\Omega$ can be
written as a set of observations $s\in \S$, each of them storing a position $
p_s$ and a (normalized) orientation $\vec{N}_s$. Poisson Surface Reconstruction
\cite{kazhdan2006poisson} aims to build a function $\rchi_{PSR}:\R^3\rightarrow
\R$ which takes positive values inside $\Omega$ and negative values outside of
it, thus making the zero levelset $\rchi_{PSR}=0$ the reconstructed surface
$\partial \Omega$.

PSR begins by building a grid $\mathcal{O}$. Then, they define the
following vector field for any $q\in\R^3$ by using the grid structure to
interpolate the orientation observations:
\begin{equation}\label{equ:vectorfield}
    \vec{V}_{PSR}(q) = \sum_{s\in \S} \frac{1}{W( p_s)} \sum_{o\in B(s)} \alpha_{o, p_s} F_o(q)\vec{N}_s 
\end{equation}
where $B(s)$ are the eight closest grid nodes to $s$, $\alpha_{o, p_s}$ is the
trilinear interpolation weight for $ p_s$ at $o$, $W$ is a measure of volumetric
sampling density and $F_o$ is a compactly-supported approximation of a Gaussian
kernel centered at the $o$-th node.

The fundamental observation of PSR is that once $\vec{V}$ has been constructed,
the desired implicit function $\rchi$ satisfies
\begin{equation}\label{equ:poisson}
    \Delta \rchi_{PSR}(x)  = \nabla \cdot \vec{V}_{PSR}(x)\,,\qquad\forall x\in\mathcal{O}
\end{equation}
\refequ{poisson} is underdetermined, as wildly different functions can have the
same Laplacian. This ambiguity is resolved in part during discretization.
\citet{kazhdan2006poisson} suggest building $\mathcal{O}$ as an adaptive grid of
a bounding box of the point cloud. Then, they use the finite element method to
build discrete Laplacian and divergence operators $\mathbf{L}$ and $\Z$ and
solve
\begin{equation}\label{equ:poissoneq}
    \mathbf{L} \mathbf{f}_{PSR} = \Z \mathbf{v}_{PSR}.
\end{equation}
This discretization imposes zero Neumann $\nabla f\cdot\vec{n} = 0$ boundary conditions on the boundary of $\mathcal{O}$. Even so, \refequ{poissoneq} only determines $\mathbf{f}$ up to translation. To account for this, one valid $\mathbf{f}_{PSR}$ is
computed and then shifted so its values near the observation points are zero on
average. The solution $\mathbf{f}_{PSR}$ contains the implicit function $\rchi_{PSR}$ evaluated at each grid node, and its zero levelset can be extracted using
contouring algorithms like Marching Cubes \cite{lorensen1987marching} or Dual
Contouring \cite{ju2002dual}.

\section{Stochastic Poisson Surface Reconstruction}\label{sec:spsr}

\begin{figure}
    \centering
    \includegraphics{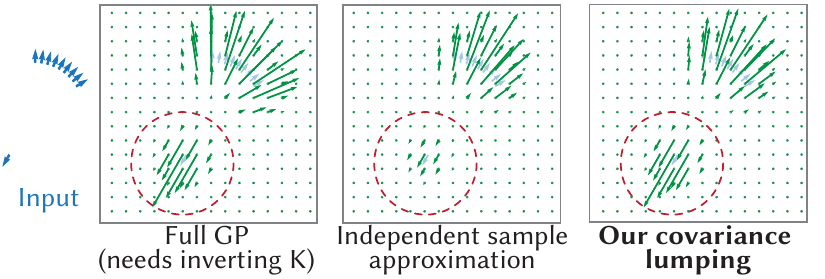}
    \caption{One can avoid the GP sample covariance matrix inversion (center left) by assuming samples to be independent (center right). This makes magnitudes proportional to sampling density (see highlight). Our \emph{covariance lumping} approximates the full GP with invariant magnitudes (right).}\label{fig:covariance-lumping}
\end{figure}

We introduce our \emph{Stochastic PSR}, which extends the traditional PSR from \citet{kazhdan2006poisson} with the statistical formalism of a Gaussian Process. We begin by
defining the \emph{PSR semicovariance} as 
\begin{equation}
    k_{PSR}(x,y) = \sigma_{g} \sum_{o\in B(x)} \alpha_{o,x} F_o(y)\, ,
\end{equation}
where $\sigma_g$ is a scalar parameter. Then, the vector field interpolation in
\refequ{vectorfield} can be written as 
\begin{equation}
    \vec{V}_{PSR}(q) = \sum_{s\in S} k_{PSR}(p_s,q) \frac{1}{\sigma_g W(p_s)}  \vec{N}_s \, .
\end{equation}

By term identification, it may be tempting to interpret this interpolation directly as the posterior mean of a Gaussian Process $\k_2^\top \K_3^{-1} \mathbf{y}$ (see \refequ{conditionalprobability}), with normals as the observed values $\mathbf{y}$, $k_{PSR}(p_s,q)$ as the entries of $\k_2$ and division by $\sigma_g W(p_s)$ playing the role of multiplication by $\K_3^{-1}$. However, a valid Gaussian distribution must have a symmetric covariance, and $k_{PSR}(p_s,q)\neq k_{PSR}(q,p_s)$ in general (hence the \emph{semi} in ``PSR \emph{semi}covariance''). To circumvent this, we define the \emph{Stochastic PSR covariance} to be its
symmetrized version
\begin{equation}\label{equ:symmetric}
    k_{SPSR}(x,y) = k_{SPSR}(y,x) = \frac{1}{2}(k_{PSR}(x,y) + k_{PSR}(y,x))\, ,
\end{equation}
which analagously defines a \emph{Stochastic PSR vector field}
\begin{equation}\label{equ:SPSRvectorfield}
    \vec{V}_{SPSR}(q) = \sum_{s\in S} k_{SPSR}(p_s,q) \frac{1}{\sigma_{g} W(p_s)}   \vec{N}_s 
\end{equation}

This variance symmetrization will be our only deviation from the traditional
Poisson Surface Reconstruction by \citet{kazhdan2006poisson} (see
\reffig{covariance-didactic}). Strictly speaking, all the observations that
follow in this paper can only be said to apply to $\vec{V}_{SPSR}$; however,
note that $\vec{V}_{PSR}$ and $\vec{V}_{SPSR}$ are visually indistinguishable
(see \reffig{vector-field-didactic}) and their difference vanishes under refinement (see proof in \refapp{convergence-proof}).

Our critical observation is that the interpolation in \refequ{SPSRvectorfield}
can be interpreted as the mean of a supervised learning task under the
assumptions of a Gaussian Process with $m=0$ and covariance $k_{SPSR}$. Indeed, one
needs only to exchange $A$ for the vector-valued $\vec{V}$ and the observations
$\{(x_i,a_i)\}$ for $\{(p_s,\vec{N}_s)\}$ in \refequ{conditionalprobability} to
obtain the conditional probability distribution
\begin{equation}
    \vec{V}(q)\,|\,\mathcal{S} \sim \N (\k_2^\top \K_3^{-1} \vec{N}_s,k_1 - \k_2^\top \K_3^{-1} \k_2)
\end{equation}
where 
\begin{align}
    k_1 = k(q,q)\,,\quad \k_2 = (k(q,p_s))_{s\in\S}\in\R^{|\S|}\,,\\ 
    \K_3 = (k(p_s,p_{s'}))_{s,s'\in\S}\in\R^{|S|\times|S|}\,.
\end{align}
The main computational cost in computing the mean and variance of the
conditional distribution is the inversion of $\K_3$, a matrix whose size scales with the number of points in the cloud. One could avoid this by assuming that the samples $\mathcal{S}$ are independent, which would result in approximating $\K_3$ by $\sigma_g \mathbf{I}$. However, completely discarding the sample interdependency could have adverse effects: in practice, it would make densely sampled regions have an outsized effect over more sparsely sampled ones (see \reffig{covariance-lumping}, center right).

We introduce a better approximation: we assume each sample is independent, but with variance proportional to sampling density. Intuitively, this means we account for the interdependence by trusting each individual sample in a densely sampled region less than in a sparsely sampled one (see \reffig{covariance-lumping}, right). We justify this choice mathematically in the general GP context in \refapp{lumping}. 

\begin{figure}
    \centering
    \includegraphics{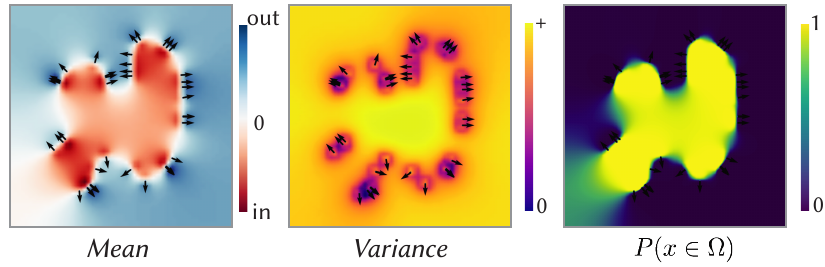}
    \vspace{-0.3cm}
    \caption{\new{By adding a variance, we can consistently compute the probability
    of any point in space being in the sampled
    object.}}\label{fig:probability-example}
\end{figure}

We are not aware of any previous work that proposes this approximation, which addresses one of the main performance limitations of general Gaussian Processes. Numerically, this resembles the mass matrix lumping step often carried out in the Finite Element literature (see e.g., \cite{zienkiewicz2005finite} Chap. 16.2.4.), so we call it the (diagonal) \emph{lumped covariance matrix} 
\begin{equation}\label{equ:covariance-diag}
    \D:=\diag(\sigma_g w) \approx \K_3
\end{equation} 
\new{where $w$ is the local sampling density at each sample point, which we compute as described by \citet{kazhdan2006poisson}}. As is usual with
Gaussian processes, we sum $\sigma_n\mathbf{I}$ if the sampled point
cloud is assumed to contain noise with variance $\sigma_n$ \new{in the normal vector}. 

Then, the
conditional distribution becomes
\begin{equation}
    \vec{V}(q)\,|\,\mathcal{S} \sim \N (\k_2^\top \D^{-1} \vec{N}_s,k_1 - \k_2^\top \D^{-1} \k_2)
\end{equation}
Importantly, note that the mean of this distribution is \emph{exactly} the
Stochastic PSR vector field we introduced in \refequ{SPSRvectorfield}, i.e.,
\begin{equation}
    \vec{V}(q)\,|\,\mathcal{S} \sim \N (\vec{V}_{SPSR}(q),k_1 - \k_2^\top \D^{-1} \k_2)\, .
\end{equation}

In other words, this critical reinterpretation of PSR has allowed us to extend
the interpolated PSR vector field into a complete statistical distribution 
\new{whose
mean $\vec{V}_{SPSR}$ is nearly identical to $\vec{V}_{PSR}$ (see \reffig{vector-field-didactic}, which we can now understand as comparing the traditional PSR vector field to the mean of our lumped Gaussian Process).} 

Ideally, however, we would want a
similar statistical formalism not for $\vec{V}$ but for the implicit function
$\rchi$, which would allow us to formulate statistical queries meaningful to the
reconstruction task, e.g., $P(\rchi(q)<0)$.

\begin{figure}
    \centering
    \includegraphics{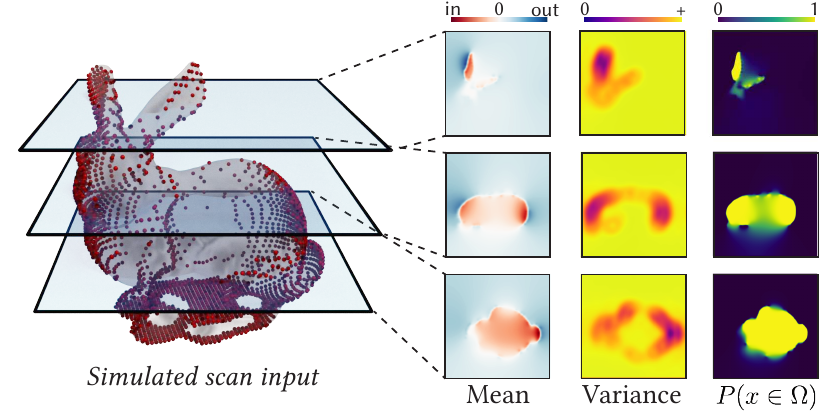}
    \vspace{-0.3cm}
    \caption{\new{Our extension of PSR provides a mean and variance, which can be
    used to compute probabilities.}}\label{fig:slice-planes}
\end{figure}

Fortunately, transfering our statistical understanding of $\vec{V}$ to $\rchi$
is just a matter of Gaussian arithmetic. First, note that a reasoning identical
to the one above leads to the joint conditional probability of $\vec{V}$ at all
grid nodes being
\begin{equation}\label{equ:allnodesdistr}
    \vec{V}(o_1),\dots,\vec{V}(o_{|O|})\,|\,\mathcal{S} \sim \N (\V_{SPSR},\K_1 - \K_2^\top \D^{-1}\K_2)\, ,
\end{equation}
where
\begin{align}
    \K_1 = (k(o,o'))\in\R^{|\mathcal{O}|\times|\mathcal{O}|}\,,\quad \K_2 = (k(o,p_s))\in\R^{|\mathcal{O}|\times |\S|}
\end{align}
and
\begin{align} 
    \V_{SPSR} = (\vec{V}_{SPSR}(o))\in\R^{|\mathcal{O}|\times 3}\,.
\end{align}
For simplicity, denote $\K_{\V} = \K_1 - \K_2^\top \D^{-1} \K_2$,
and let $\mathbf{v}$ be the concatenated vector of $\vec{V}$ values as in the
previous section. Then, \refequ{allnodesdistr} becomes
\begin{equation}
    \mathbf{v}\,|\,\mathcal{S} \sim \N (\V_{SPSR},\K_{\V})\,.
\end{equation}
By linearity and bilinearity of the mean and covariance, respectively, we know
\begin{equation}
    \Z\mathbf{v}\,|\,\mathcal{S} \sim \N (\Z\V_{SPSR},\Z\K_{\V}\Z^\top)\,.
\end{equation}
and thus, since the vector $\mathbf{f}$ of evaluations of $\rchi$ satisfies
$\L\mathbf{f}=\Z\mathbf{v}$,
\begin{equation}
    \mathbf{f}\,|\,\mathcal{S}\sim \N \left(\mathbf{L}^{-1}\Z\V_{SPSR},\mathbf{L}^{-1}\Z\K_{\V}\Z^\top(\mathbf{L}^{\top})^{-1}\right)\,,
\end{equation}
Similarly to the traditional PSR, we shift the mean values to be zero near the
sample points and shift the variance values to have a minimum diagonal entry of zero. We write the above distribution as
\begin{equation}\label{equ:spsr}
    \mathbf{f}\,|\,\mathcal{S}\sim \N \left(\mathbf{f}_{SPSR},\K_{\mathbf{f}}\right)\,.
\end{equation}
This statement is our principal contribution, as it extends the traditional PSR
implicit function $\mathbf{f}_{PSR}$ into a full statistical distribution
whose mean is almost identical to $\mathbf{f}_{PSR}$. Of all the information
contained in $\K_{\mathbf{f}}$, a particularly useful quantity is its diagonal
$\sigma^2_{\mathbf{f}}$, which contains the variances of each entry of $f$
(see \reffig{psr-vs-spsr}). 

\subsection{Statistical queries}\label{sec:queries}

\begin{figure}
    \centering
    \includegraphics{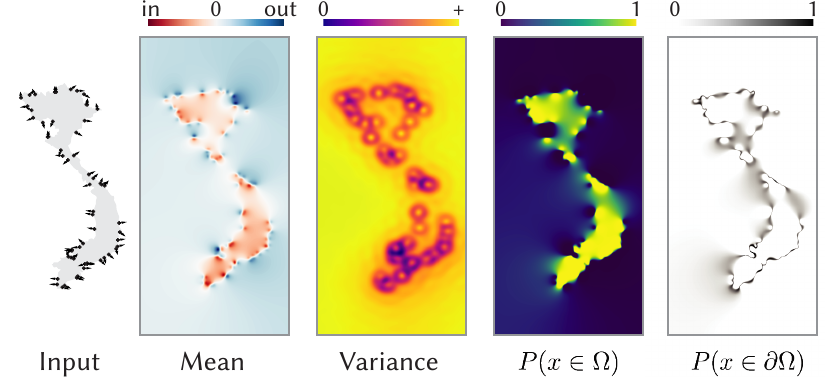}
    \vspace{-0.3cm}
    \caption{\new{Our computed mean and variance (center left) completely determine
    the implicit function's distribution, allowing us to respond to statistical
    queries like the likelihood of a point being in the volume defined by the
    point cloud (center right) or on its surface
    (right).}}\label{fig:surface-probability-example}
\end{figure}

\begin{figure}
    \centering
    \includegraphics{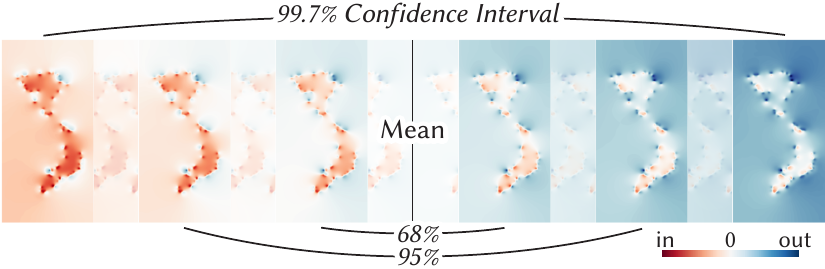}
    \vspace{-0.3cm}
    \caption{\new{The \emph{68-95-99.7 rule} from normal distributions lets us
    compute confidence intervals for the implicit function values. Regions with
    the same color (red or blue) at both extremes of an interval can be said to
    be inside or outside with at least the interval's
    confidence.}}\label{fig:confidence-interval}
\end{figure}

\refequ{spsr} contains all the
information we need to respond to statistical queries fundamental to the
reconstruction process; e.g., the probability of a given point being
inside the shape (see \reffig{probability-example})
\begin{equation}
    p(o_i\in\Omega) = p(\mathbf{f}_i\leq 0) = CDF_{\mathbf{f}_{SPSR,i},\sigma^2_i} (0)
\end{equation}
or the probability density of being on the surface (see
\reffig{surface-probability-example})
\begin{equation}
    p(o_i\in\Omega) = p(\mathbf{f}_i=0) = PDF_{\mathbf{f}_{SPSR,i},\sigma^2_i} (0)
\end{equation}
where $CDF_{\mu,\sigma^2}$ and $PDF_{\mu,\sigma^2}$ are the cummulative distribution
and probability density functions, respectively, of a Gaussian distribution with
mean $\mu$ and variance $\sigma^2$. Similar Gaussian arithmetic lets us compute
other quantities like confidence intervals (see \reffig{confidence-interval}).

\begin{figure}
    \centering
    \includegraphics{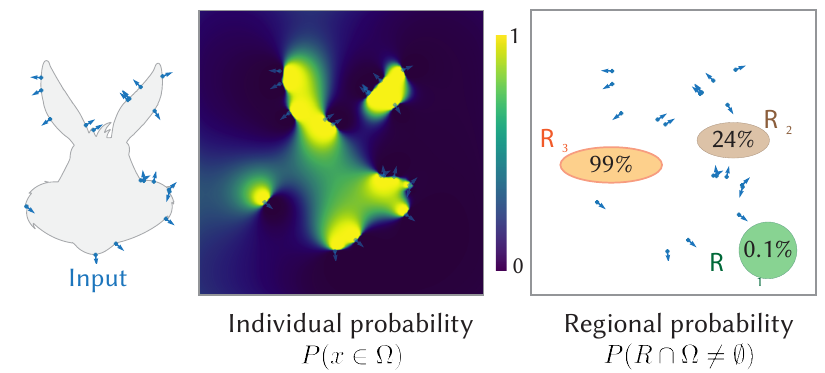}
    \vspace{-0.3cm}
    \caption{\new{Our algorithm not only allows us to compute individual spatial probabilities (center), but also joint regional probabilities (right).}}\label{fig:integrate-region-full}
\end{figure}

All these queries would be impossible to answer from PSR's 
$f_{PSR}$ alone. We show this in the didactic example in \reffig{convergence}, where
vectors pointing upward are sampled from a normal distribution centered at the
middle horizontal line. As more samples get added, $f_{PSR}$ converges to a
smooth transition between positive and negative values; however, our variances
are progressively decreasing, making $P(\x\in\Omega)$ converge in certainty as
expected.

We formalize this observation further by defining an integrated statistical
quantity, the \emph{total uncertainty} $U_{SPSR}$, which we define as 
\begin{equation}
    U_{SPSR} = \int_{B} \left( 0.5 - |P(\x\in\Omega) - 0.5| \right) \,dx
\end{equation}
where $B$ is a bounding box around the point cloud. $U_{SPSR}$ can be
intuitively interpreted as a global measure of reconstruction quality which
converges to zero when the surface has been reconstructed with absolute
certainty (see \reffig{total-uncertainty}). 
\new{We pose that this quantity
can
be used as a stopping criteria to guide the scanning process (see
\reffig{total-uncertainty-threshold}).
We know of no analogous quantity that can be formally defined from the traditional PSR understanding alone.}

The off-diagonal entries of $\K_\mathbf{f}$ quantify the correlation between the values of $f$ at different points in $\mathcal{O}$. This allows for \emph{joint probability queries}; for example, computing the likelihood of a whole region in space $R$ intersecting $\Omega$ (see \reffig{integrate-region-full}), a quantity of immediate relevance to collision detection applications. 

To compute it, we can randomly sample $R_s=\{r_1,\dots,r_s\}\subset R$. If $\mathbf{W}$ is the linear interpolation matrix from $\mathcal{O}$ to $R_s$, then distribution arithmetic says
\begin{equation}\label{equ:regionprob}
    \mathbf{f}(r_1,\dots,r_s)\, | \mathcal{S} \sim \mathcal{N}(\mathbf{W}\mathbf{f}_{SPSR},\mathbf{W}^\top\K_{\mathbf{f}}\mathbf{W})\,.
\end{equation}  
The collision probability, i.e., the probablity that any $r_i$ is contained in the shape, is opposite to the joint probability that all $r_i$ are outside the shape, $p(\mathbf{f}(r_1)>0,\dots,\mathbf{f}(r_s)>0))$. This quantity is also the same as $p(-\mathbf{f}(r_1)<0,\dots,-\mathbf{f}(r_s)<0))$, known as the \emph{multivariate Gaussian cummulative distribution} of $-\mathbf{f}$, which can be estimated with numerical integration (see Figs.~\ref{fig:integrate-region-full} and \ref{fig:car-collision}).

\subsection{Space reduction with eigenspace analysis}

All this additional information comes at a computational cost. Specifically, the
main performance bottleneck is the computation of the covariance matrix 
\begin{equation}
  \K_{\mathbf{f}}  = \mathbf{L}^{-1}\Z\K_{\V}\Z^\top(\mathbf{L}^{\top})^{-1}\, ,
\end{equation}
a matrix equation that amounts to solving $2\times|\mathcal{O}|$ linear systems.
Even making use of efficient prefactorizations of $\L$, this step is
impracticable for large 3D grids with sizes in the millions of cells. 

To circunvent this, we will work in the reduced space of Laplacian eigenvectors. For a cuboid with lengths $\ell_1,\ell_2,\ell_3$ and zero Neumann boundary conditions, these are known \cite{gottlieb1985eigenvalues} to respond to the analytic function
\begin{equation}
    \psi_{M,N,\tilde{N}}(x,y,z) = \cos \left( \frac{M\pi x}{\ell_1}\right)\cos\left(\frac{N\pi y}{\ell_2}\right) \cos \left( \frac{\tilde{N}\pi z}{\ell_3}\right)\,,
\end{equation}
with associated eigenvalues
\begin{equation}
    \lambda_{M,N,\tilde{N}} = \pi^2\left[ \left(\frac{M}{\ell_1}\right)^2 + \left(\frac{N}{\ell_2}\right)^2 + \left(\frac{\tilde{N}}{\ell_3}\right)^2  \right]\,.
\end{equation}

Specifically, we will evalute these analytical expressions into $\E\in\R^{|\mathcal{O}|\times k}$, the matrix
of the $k$ lowest-magnitude eigenvectors of $\L$, and $\D_{e}$, the
diagonal eigenvalue matrix. Then, we approximate $\K_{\mathbf{f}}$ by projecting
$\Z\K_{\V}\Z^{\top}$ to the eigenspace, solving the matrix equation in this
reduced space, and reprojecting; i.e., 
\begin{equation}\label{equ:projection}
    \K_{\mathbf{f}}  \approx \E \D_{e}^{-1} \E^{\top} \left(\Z \K_{\V}\Z^\top \right) \E \D_{e}^{-1} \E^{\top}\, .
\end{equation}
In all our examples, we fix $k=3000$. 

Due to the dimensionality of the above matrices, the mostly computationally expensive step in \refequ{projection} is the product $\E\C\E^{\top}$, where $\C\in\R^{k\times k}$ results from multiplying all middle matrices in \refequ{projection}. We avoid this step by noting that it is rare that one is interested in all entries of $\K_\mathbf{f}$. As seen in \refsec{queries}, most statistical queries require only the diagonal $\sigma^2_{\mathbf{f}}$, which we can compute directly and efficiently as $(\E\C)\cdot \E$, where $\cdot$ denotes row-wise dot product. In the case of joint probability queries that require off-diagonal knowledge of $\K_{\mathbf{f}}$, we first construct the selection matrices $\mathbf{S}, \mathbf{S}^\top$ that extract the covariance rows and columns relevant to the specific application, and directly compute $\E'\C\E'^{\top}$, where $\E' = \mathbf{S}\E$. Thus, we eventually avoid ever computing (or storing) the full $\K_{\mathbf{f}}\in\R^{|\mathcal{O}|\times|\mathcal{O}|}$.

\new{Dimension reduction has been used for implicit reconstruction before; e.g., the Fourier coefficients computed by \citet{kazhdan2005reconstruction}. We limit our approximation to the covariance computation only.}

\subsection{Adaptive discretization}\label{sec:adaptive}

\begin{figure}
    \centering
    \includegraphics{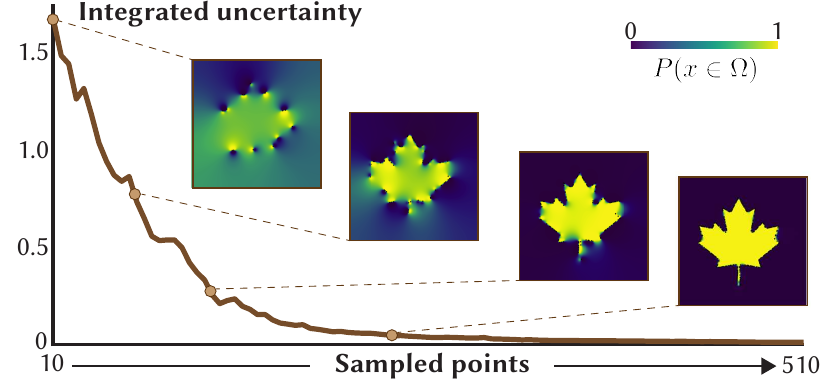}
    \vspace{-0.3cm}
    \caption{\new{Our statistical formalism means we can define quantities like the
    integrated uncertainty, which converges to zero as the probability is
    collapsed to zero and one. This measure can serve as a scanning stopping
    criterion.}}\label{fig:total-uncertainty}
\end{figure}

\begin{figure}
    \centering
    \includegraphics{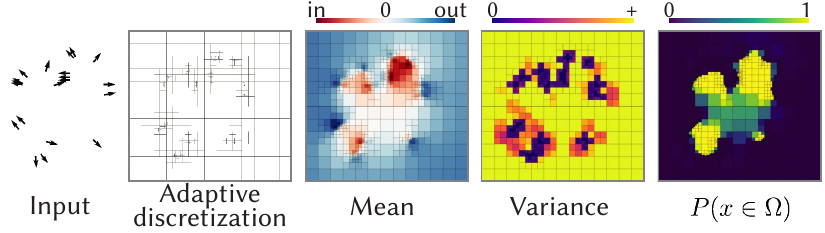}
    \vspace{-0.3cm}
    \caption{\new{Our contribution is orthogonal to the choice of discretization, as we show by answering statistical queries on a graded quadtree.}}\label{fig:quadtree}
\end{figure}

\begin{figure}
    \centering
    \includegraphics{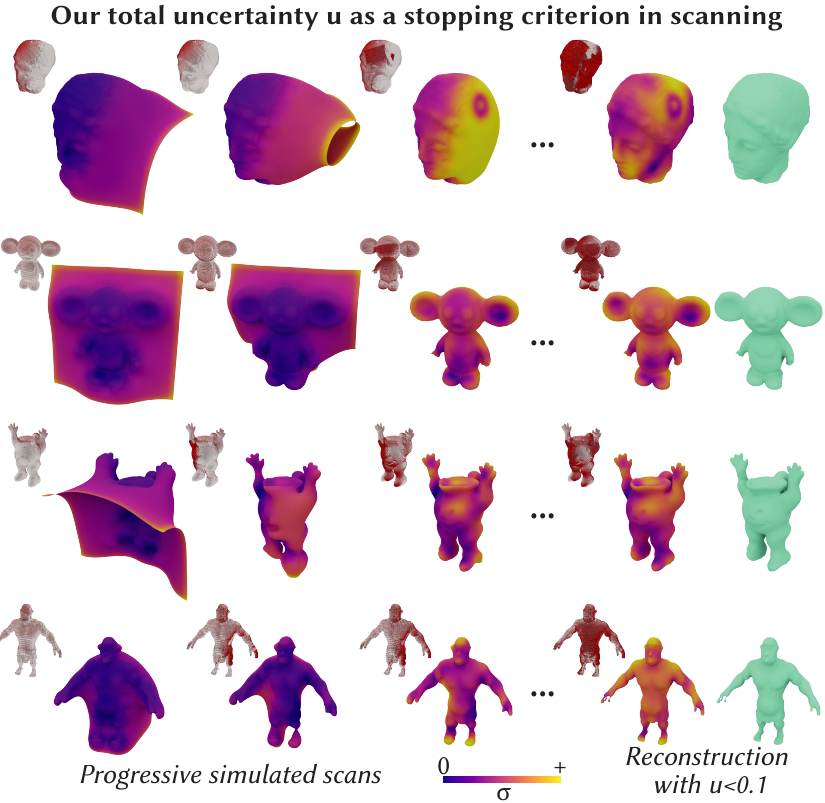}
    \vspace{-0.3cm}
    \caption{\new{Like PSR, we can extract isosurfaces of the mean of our Stochastic
    PSR. Unlike PSR, our statistical formalism provides a reliable
    stopping criterion by thresholding integrated uncertainty $u$. Meshes
    colored by variance.}}\label{fig:total-uncertainty-threshold}
\end{figure}

\begin{figure*}
    \centering
    \includegraphics{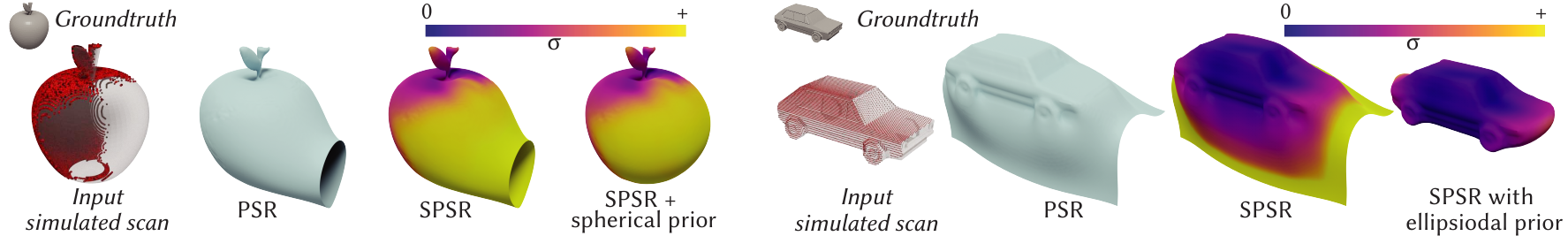}
    \vspace{-0.6cm}
    \caption{\new{Taking inspiration from Figs. 6 and 7 in
    \cite{martens2016geometric}, we show the result of reconstructing an apple
    and a car from a partial scan. PSR fails to produce even a closed surface,
    as does our vanilla SPSR, albeit also providing variance information
    signaling the less confident regions. When combined with a task-specific
    simple primitive (spherical or ellipsoidal) prior, SPSR provides a
    significantly better reconstruction.}}\label{fig:geometric-priors-examples}
\end{figure*}

The statistical formalism we contribute is agnostic to where the values of $f$ and $\vec{V}$ are stored and to the discretization scheme used to solve the Poisson equation. \citet{kazhdan2006poisson} suggest building an adaptive grid which is finer near the sampled point cloud. 
This choice is justified through their sole goal of accurately recovering the zero levelset of $f$.

Our Stochastic PSR extends the traditional PSR by providing volumetric
statistical information, which is also relevant and non-trivial in regions away
from the zero levelset of our mean function $\mathbf{f}_{SPSR}$ or the point
cloud (see, e.g., the probability in \reffig{probability-example}). Further,
computing statistical queries accurately away from the mean-zero levelset can be
critical for applications that our formalism newly allows like collision
detection or ray casting (see \reffig{ray-casting}). Thus, the choice of an
adaptive grid structure is less obvious for our algorithm. Nonetheless, we show
in \reffig{quadtree} our algorithm working as expected on a graded quadtree grid
structure, following the finite difference discretization proposed by
\citet{bickel2006adaptive}. While orthogonal to our contribution, we believe
exploring different adaptive discretization strategies (e.g., narrowing in on
regions of higher \emph{uncertainty}) to be a promising avenue for future work.

\section{Beyond Poisson Surface Reconstruction}\label{sec:beyondspsr}

A significant benefit of our interpretation of Poisson Surface Reconstruction as
a Gaussian Process is that we may borrow from the vast literature on the latter
to study variations of the traditional PSR. A prototypical example of this is
incorporating task-specific priors.

In \refsec{gp}, we mentioned that most Gaussian Process consider $m=0$ and are
thus fully determined by the covariance $k$ (this is often known as
\emph{kriging} in the GP literature). Indeed, it was by assuming $m=0$ that we
recovered the PSR vector field interpolation as the mean of the GP posterior
distribution. In fact, the choice of $m$ is a prior imposed on the joint
distribution before any data is observed. Thus, PSR can be understood
to have a zero-gradient prior.

Incidentally, this observation also provides an answer to an often asked
question about PSR; namely, why PSR, which is posed as recovering a function
which is zero on the surface and satisfies certain norm-one gradients, produces
an output so different from a Signed Distance Field (SDF). From this novel
Gaussian Process perspective, we can understand that PSR is, via its $m=0$
prior, encouraging zero-norm gradients away from the sample positions, while an
SDF requires Eikonal norm one gradients almost everywhere.

\paragraph{Primitive geometric priors.}
\citet{martens2016geometric} suggest using simple geometric primitives as priors
in the context of GP implicit surfaces. Since the authors are directly learning the
surface's implicit representation, their suggested priors $m$ are scalar,
SDF-like functions. In our case, where learning is carried out in the gradient
space and then transfered to an implicit function via a PDE solve, we can use
gradient vector-valued priors.

\begin{figure}
    \centering
    \includegraphics{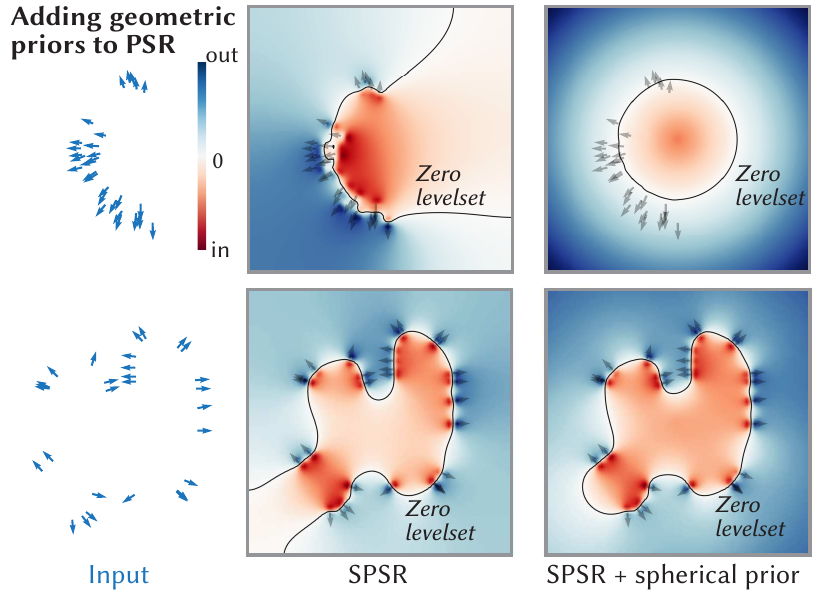}
    \vspace{-0.3cm}
    \caption{\new{Adding a spherical prior to our Stochastic PSR reconstruction helps
    us recover spherical objects and also avoid open surfaces in
    partial scans.}}\label{fig:spherical-prior-didactic}
\end{figure}

In \reffig{spherical-prior-didactic}, we exemplify the effect of a simple
spherical prior
\new{
\begin{equation}
    m(\x) = \alpha \frac{\x-\mathbf{c}}{\|x-\mathbf{c}\|}\, ,
\end{equation}}
which we pose can be a useful tool for favouring closed reconstructions over
open ones.
\new{$\alpha$ is a parameter tuning the strength of the prior
and $\mathbf{c}$ is the center of the sphere, which we set to be the average of all point cloud positions.}
In
\reffig{geometric-priors-examples}, we show less didactical uses of
this same prior, along with a similar ellipsoidal one.

\section{Implementation details}

We implemented our algorithm in \textsc{Python}, using \textsc{libigl}
\cite{libigl} for common geometry processing subroutines. We rendered all our
figures in \textsc{Blender}, using \textsc{BlenderToolbox}
\cite{blendertoolbox}.  All our results were produced on our machine with Intel Xeon CPU E5-2637 v3 \@ 3.50Hz (16 cores) with 64GB of RAM.

For parameter standardization, we normalized all our examples to fit a
length-one cube as a preprocessing step. Unless specified otherwise, we use a
$100^3$ uniform grid ($100^2$ in the two-dimensional examples) such that $\L$ and $\Z$ become the usual finite difference Laplacian and divergence matrices. We fix
$\sigma_g=0.02$, $\alpha=0.05$ and $k=3000$. We use the same function $F_o$ as \citet{kazhdan2006poisson}, obtained by convolving a
box filter with itself three times. 

For memory efficiency, we use SciPy's biconjugate gradient method to solve for the
distribution mean $f_{SPSR}$, which accounts for approximately $5\%$ of our
runtime (around $10$ seconds in all 3D examples). Our main computational
bottleneck is \refequ{projection} (specifically, the projection of $\K_\V$ into
the reduced space), covering $90\%$ of our runtime (around two minutes in all 3D
examples).

\begin{figure*}
    \centering
    \includegraphics{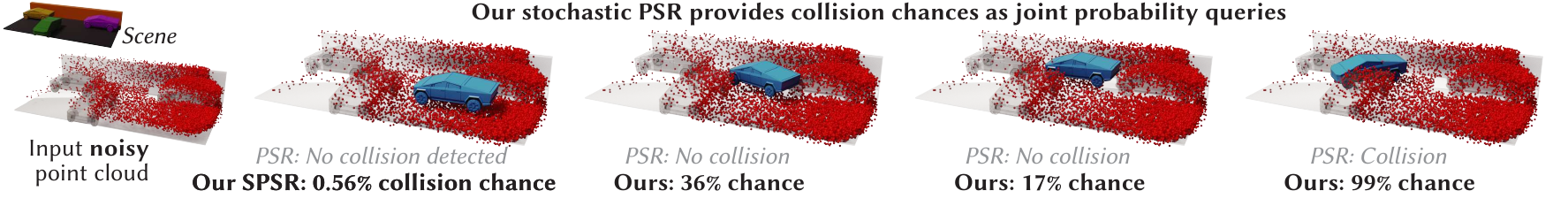}
    \vspace{-0.8cm}
    \caption{\new{We can use our statistical formalism it to compute statistical quantities like collision chances (bold), unlike traditional PSR (grey).}}\label{fig:car-collision}
    \vspace{-0.2cm}
\end{figure*}

\begin{figure}
    \centering
    \includegraphics{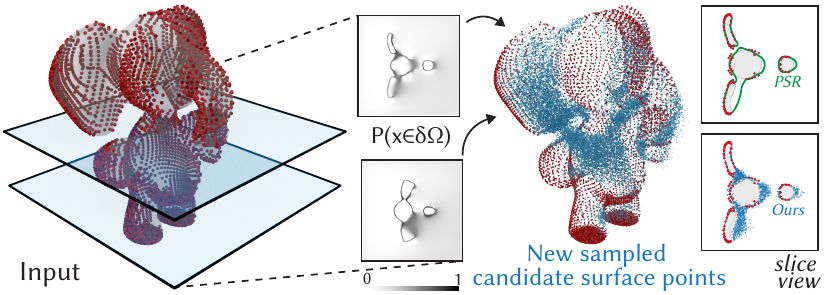}
    \vspace{-0.3cm}
    \caption{\new{We can use our computed surface likelihood to repair the input
    point cloud, naturally drawing points from less certain regions.}}\label{fig:surface-likelihood}
\end{figure}

\section{Experiments}

\subsection{Deviation from \citet{kazhdan2006poisson}} Our sole deviation from
traditional Poisson Surface Reconstruction as introduced by
\citet{kazhdan2006poisson} is the symmetrization of the covariance in
\refequ{symmetric}. Theoretically, it is clear that both $k_{PSR}$ and $k_{SPSR}$
converge to the same symmetric $F_{x}(y)$ in the limit of grid refinement.
However, we also justify the step by showing that the difference between these
two is small even at a very coarse resolution in \reffig{covariance-didactic}.
In PSR, $k_{PSR}$ is used to reconstruct a vector field $\vec{V}_{PSR}$.
In \reffig{vector-field-didactic}, we show that the interpolated vector field
$\vec{V}_{SPSR}$ obtained with the symmetrized covariance is visually identical.
In \reffig{psr-vs-spsr}, we show that this applies also to the implicit function
$\rchi$ obtained through the Poisson solve.

\subsection{Statistical quantities}
Our Stochastic PSR exchanges the functional PSR output for a fully defined
statistical distribution. To explore just how much more information is contained
in this distribution, we show different statistical quantities meaningful to the
shape reconstruction process. 

In \reffig{probability-example}, we use our computed mean and variance to
evaluate each point's cummulative distribution functions at zero, which returns
the probability of any point in space being contained in $\Omega$. Similarly,
evaluating the probabilistic density function at zero returns a measure of
surface probability, as we show in \reffig{surface-probability-example}. We use
slice planes to visualize these functions for a 3D reconstruction problem in
\reffig{teaser} and \reffig{slice-planes}. In \reffig{confidence-interval}, we
show our statistical formalism applied to the computation of confidence
intervals, which serve as an intuitive tool for visualizing variances.

\section{Applications}

Our main contribution is a novel theoretical understanding of Poisson Surface
Reconstruction that expands it beyond its traditional application realm. We
explore this with prototypical results.

\subsection{Collision detection}

Statistical information is critical for robots or autonomous vehicles, which may
capture their surroundings as point clouds with partial occlusions.
When a car is deciding whether to swerve, it does not want to know only
if the maneuvre is secure \emph{for the most likely scenario of its
surroundings} (the traditional PSR output or the SPSR mean); rather, it must
account for uncertainty and conclude whether the maneuvre is safe
\emph{in the vast majority of possible scenarios}.

\begin{figure}
    \centering
    \includegraphics{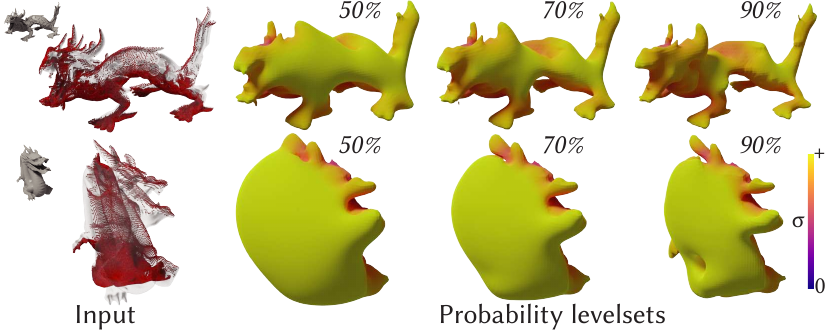}
    \vspace{-0.3cm}
    \caption{\new{We can threshold our computed probability $p(x\in\Omega)$ to obtain
    meshes that can be used for, e.g., collision
    detection.}}\label{fig:levelsets}
\end{figure}

We exemplify this in \reffig{car-collision}, where we simulate a 3D scan of a street and use joint probability queries to compute the collision chance of an automated vehicle along a given trajectory. 
\new{We compare this chance against a baseline (in grey), which only checks for intersections between the car mesh and the traditional PSR zero isolevel extracted with Marching Cubes \cite{lorensen1987marching}.}
\new{We include synthetic Gaussian noise both on the input point positions and normal vectors.}
In \reffig{levelsets}, we show how one can threshold our computed probabilities $p(x\in\Omega)$ to obtain isosurfaces for use in collision detection applications.

\subsection{Volume rendering}

\begin{figure}
    \centering
    \includegraphics{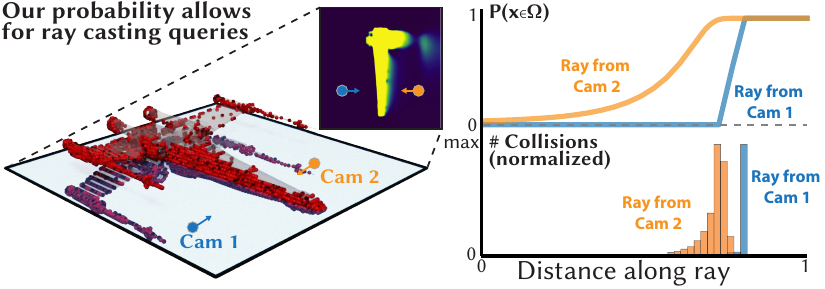}
    \vspace{-0.3cm}
    \caption{Our probability can be interpreted as a density for ray casting applications that have no analogue in the traditional PSR formulation.}\label{fig:ray-casting} 
    \vspace{-0.2cm}
\end{figure}

The arbitrary units in the traditional PSR output mean that it cannot be
interpreted as a volume rendering occupancy. 
Fortunately, our Stochastic Poisson Surface Reconstruction allows us to compute
$p(\x\in\Omega)$, values between 0 and 1 which can be interpreted
as occupancies (this reinterpretation is common in the Neural Radiance Field
literature, e.g., \cite{mildenhall2020nerf}). In \reffig{ray-casting}, we use
standard volume rendering techniques (see \cite{pharr2016physically}, Chap. 11)
to sample the free path of a ray aimed at a partial reconstruction of a Space
Wizard vehicle from two different directions. As expected, rays intersecting
regions with higher uncertainty produce a wider spread of paths\new{, as evidenced by the orange and blue histograms.}

\subsection{Reconstruction}

\begin{figure*}
    \centering
    \includegraphics{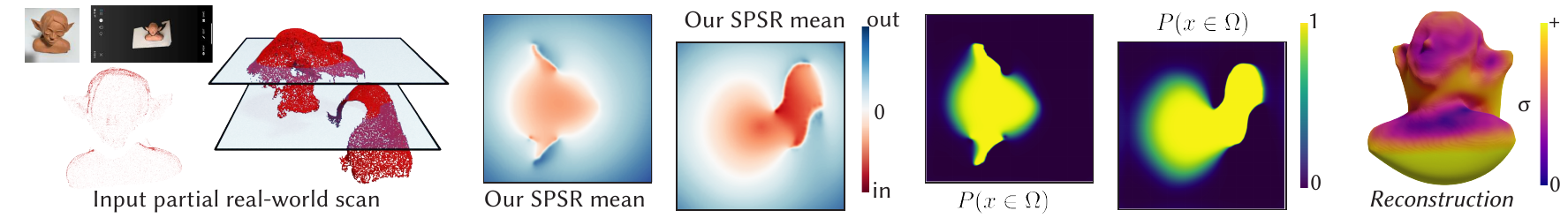}
    \vspace{-0.7cm}
    \caption{\new{Our SPSR on real-world scan data. The geometry of our reconstructed surface is the same as PSR, but we provide variances and volumetric quantities.}}\label{fig:real-scan}
    \vspace{-0.2cm}
\end{figure*}

\paragraph{Scanning integration}
Surface reconstruction is the flagship application of PSR. Despite this, PSR
cannot be fully integrated into the scanning pipeline, as it fails to provide
basic feedback like when a point cloud is dense enough or where to scan next. In
\reffig{total-uncertainty}, we show that our introduced \emph{total uncertainty} $u$ converges to zero as points are added, giving a clear measure of scan
quality. We show how one can use thresholds on $u$ as a scanning
stopping criteria in \reffig{total-uncertainty-threshold}, obtaining reconstructions of similar quality for the same threshold value. In \reffig{real-scan}, we show our algorithm's output on a real-world depth scan obtained with a smartphone app. 

The ability to simulate ray casting on our reconstructed shapes also allows us
to score different potential scan positions, as we show in
Figs.~\ref{fig:simulated-scan} and \ref{fig:camera-score}. We begin by
simulating a scan on a given groundtruth object using traditional ray tracing in
a cone around some predetermined camera positions (see \reffig{simulated-scan})
placed on the left side of a race car. 
\new{We include synthetic Gaussian noise both on the input point positions and normal vectors.}
This lets us obtain a point cloud $P$
which we can input to our algorithm and compute its total uncertainty
$U_{SPSR}(P)$. For each potential new scan position, we cast a ray and add the
resulting intersection $p$ as a new point to the input point cloud, defining the
\emph{camera score} as the change in total uncertainty from adding the new point
(see \reffig{camera-score}),
\begin{equation}
    s = |U_{SPSR}(P) - U_{SPSR}(P\cup p)|
\end{equation}

We repeat this process ten times for each potential camera position, to account
for the statistical variabilty in the ray casting. A fundamental part of any
scan feedback pipeline, this process helps us decide on a next best camera view.
As expected, scores are highest for hypothetical cameras on the right side of the
car.

\paragraph{Point cloud repair} 
Converting to an implicit function or a triangular mesh is not needed for some
downstream applications that can work directly on point clouds. However, raw
scan clouds often contain holes due to occlusions or missing camera angles. In
\reffig{surface-likelihood}, we show how our algorithm can be used 
to produce possible surface points in the occluded regions by using a Metropolis-Hastings scheme to sample from
our surface probability $p(\x\in\Omega)$.

\paragraph{Incorporating priors} 
Our novel statistical interpretation of PSR allows us to build on it even as it
applies to its most direct purpose of recovering surfaces from point clouds; for
example, by incorporating geometric priors as we exemplify in
\reffig{spherical-prior-didactic}. In \reffig{geometric-priors-examples}, we
show how even very simple primitive geometric priors can be useful for
recovering complex three-dimensional geometry.

\section{Limitations \& Conclusion}

Our work merges the renowned Poisson Surface Reconstruction with the field of Gaussian Process Implicit Surfaces, both providing a statistical formalism for PSR and a novel GPIS that can be used for supervised learning of implicit functions.

By using covariance functions with compact support like the original PSR, we sample the interpolated vector field $\vec{V}$ efficiently. However, unlike other GPIS, we require a global Poisson solve to evaluate even a single implicit function $\rchi$ mean or variance query. When seen as a GPIS, this is a limitation that makes our algorithm less efficient for applications that require a limited amount of samples. Also like the original PSR, our algorithm requires an oriented point cloud as input, and extending it to unoriented point clouds (e.g., those representing thin sheets \cite{chi2021garmentnets}) would require orienting them as a preprocessing step (see, e.g., \cite{hornung2006robust,alliez2007voronoi,metzer2021orienting}).

While computing the mean of our distribution $\x_{SPSR}$ is just as computationally expensive as the original PSR output, our covariance matrix computation requires solving a full matrix equation. We alleviate this by precomputing the Laplacian eigenpairs on a bounding box of any input. This introduces a trade-off between memory, runtime and precision when computing this covariance.

Our algorithm extends the traditional PSR by \citet{kazhdan2006poisson} by separating it into a GP vector field reconstruction and a PDE solve. It is unclear how our approach could be applied to the \emph{Screened PSR} by \citet{kazhdan2013screened}, which combines both steps to improve robustness in unsampled regions. We conjecture that the answer may lie in works on incorporating gradient observations to Gaussian processes (see 9.4 in \cite{williams2006gaussianbook}).

\begin{figure}
    \centering
    \includegraphics{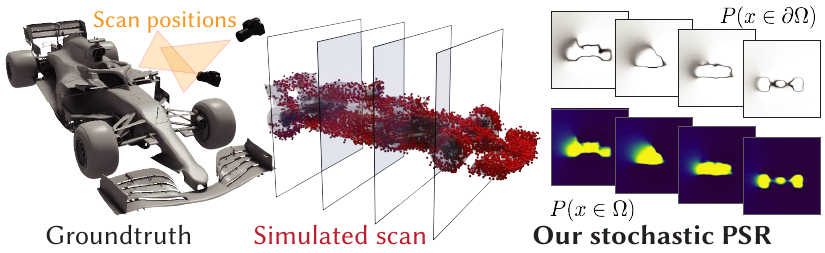}
    \vspace{-0.3cm}
    \caption{Many of our results follow a similar pipeline: we use a groundtruth object (left) and simulate scanning it from different directions to obtain an oriented point cloud (middle), which we then pass as input to our Stochastic PSR algorithm (right) to compute the desired statistical quantity.}\label{fig:simulated-scan}
\end{figure}

\begin{figure}
    \centering
    \includegraphics{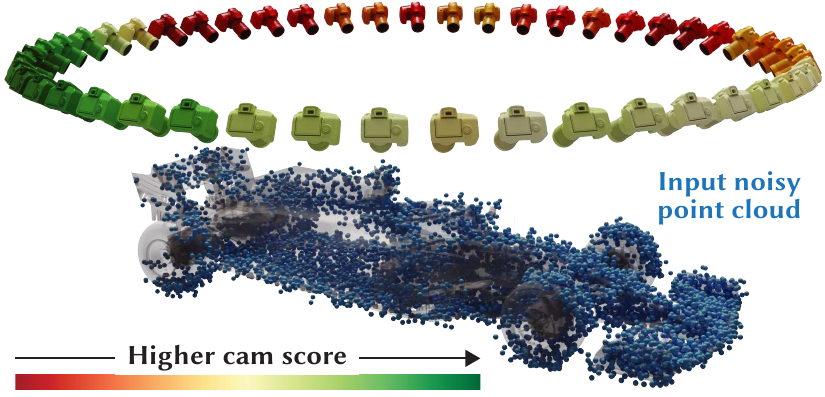}
    \vspace{-0.3cm}
    \caption{Interpreting our probability as a density and simulating rays from different cameras, we can score them according to the simulated change in total uncertainty to aid next view planning.}\label{fig:camera-score}
\end{figure}

Our work not only extends our knowledge of Poisson Surface Reconstruction from the perspective of a Gaussian Process. Indeed, we have also used PSR to expand our knowledge of Gaussian Processes, by introducing a covariance lumping technique which avoids costly matrix inversions at test time. We believe applying this in the broader GP context to be a promising avenue for future work.

As 3D acquisition becomes more accessible and attractive to consumers and the lines between real and virtual geometry get blurred, we believe quantifying and transmitting the uncertainties of the capture process to be one of the fundamental problems of this era. We hope our work inspires our geometry processing colleagues, whose workflow often involves PSR, to study how this statistical formalism carries over further down the shape processing pipeline.

\begin{acks}
This project is funded in part by NSERC Discovery (RGPIN2017–05235, RGPAS–2017–507938), New Frontiers of Research Fund (NFRFE–201), the Ontario Early Research Award program, the Canada Research Chairs Program, a Sloan Research Fellowship, the DSI Catalyst Grant program and gifts by Adobe Inc. The first author is funded in part by an NSERC Vanier Scholarship and an Adobe Research Fellowship. 

We acknowledge the authors of the 3D models used throughout this paper and thank
them for making them available for academic use: ShaggyDude (\reffig{teaser}, CC BY 4.0), Perry Engel (\reffig{total-uncertainty-threshold} third row, CC BY-NC 4.0), Joshua Poh (\reffig{geometric-priors-examples} left, CC BY-SA 3.0), Agustin Flowalistik (\reffig{geometric-priors-examples} right, CC BY-NC-SA 4.0), Will McKay (\reffig{car-collision}, CC BY-NC 4.0), Karl (\reffig{ray-casting}, CC BY-SA 3.0) and Rafael Rodrigues (Figs.~\ref{fig:simulated-scan} and \ref{fig:camera-score}, CC BY-NC-SA 4.0).

We thank Kirill Serkh, Kiriakos Kutulakos, Eitan Grinspun, David I.W. Levin, Oded Stein, Nicholas Sharp, Otman Benchekroun and Towaki Takikawa for insightful conversations that inspired and guided us throughout our work; Abhishek Madan, Aravind Ramakrishnan and Jonathan Panuelos for proofreading; Hsueh-Ti Derek Liu for help rendering our results; Xuan Dam, John Hancock and all the University of Toronto Department of Computer Science research, administrative and maintenance staff that literally kept our lab running during very hard years.  
\end{acks}

\bibliographystyle{ACM-Reference-Format}
\bibliography{references.bib}

\appendix
\section{SPSR covariance convergence proof}\label{app:convergence-proof}

We will prove that the difference between the PSR semicovariance
\begin{equation}
    k_{PSR}(x,y) = \sigma_{g} \sum_{o\in N(x)} \alpha_{o,x} F_o(y)
\end{equation}
and our Stochastic PSR covariance 
\begin{equation}\label{equ:symmetric}
    k_{SPSR}(x,y) = k_{SPSR}(y,x) = \frac{1}{2}(k_{PSR}(x,y) + k_{PSR}(y,x))\, ,
\end{equation}
vanishes at the limit of grid refinement. We can do so by showing that $k_{PSR}(x,y)$ converges to the gridless covariance function
\begin{equation}
    k^\star (x,y) = \sigma_g F_y(x)
\end{equation}
Let $\lambda$ be the Lipschitz constant of $F$. Then, since $F$ is symmetric,
\begin{equation}
    |F_o(y) - F_y(x)| = |F_y(o) - F_y(x)| \leq \lambda \| o - x\|
\end{equation}
Since trilinear interpolation weights sum up to one, this means
\begin{equation}
    |k_{PSR}(x,y) - k^\star (x,y)|\leq \sigma_g \lambda \| o - x\|\,,
\end{equation}
Symmetrically, we get 
\begin{equation}
    |k_{PSR}(y,x) - k^\star (x,y)|\leq \sigma_g \lambda \| o - x\|\,,
\end{equation}
which, by definition of $k_{SPSR}$, means
\begin{equation}
    |k_{SPSR}(x,y) - k_{PSR}(x,y)|\leq \sigma_g \lambda \| o - x\|\,.\qed
\end{equation}

\section{Justification for covariance lumping}\label{app:lumping}

\begin{figure}
    \centering
    \includegraphics{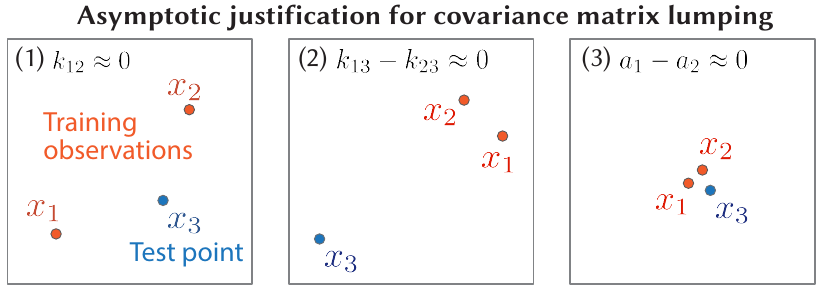}
    \vspace{-0.3cm}
    \caption{Cases where our lumping has correct asymptotics: (1) when training data are far from one another, (2) when test points are far from training data and (3) when training data are so close that their features are similar.}\label{fig:lumped-covariance-didactic}
\end{figure}

Consider a simple Gaussian Process $\mathcal{A}=\{A(\x)\}_{\x\in D}$ with zero mean and covariance function $k:D\times D\rightarrow\R$, with two different training observations $\{(x_1,a_1),(x_2,a_2)\}$ and one test point $x_3$. For clarity, let $k_{ij}=k(x_i,x_j)$. Then, since $k_{11} = k_{22}, \,k_{12} = k_{21}$, we have
\begin{equation}
    \K_3 = \begin{pmatrix}
        k_{11}   & k_{12}     \\
        k_{12} & k_{11}
    \end{pmatrix}\quad , \quad
    \k_2 = \begin{pmatrix}
        k_{13}  \\
        k_{23} 
    \end{pmatrix}
\end{equation}
Traditionally, one uses these matrices to compute the GP posterior mean $\k_2^\top\K^{-1}_3\a$ and covariance $\k_2^\top\K^{-1}_3\k_2$. Let us write both of these as $\k_2^\top \K_3^{-1}\mathbf{c}$ for some generic feature vector $\mathbf{c}$. Then,  
\begin{equation*}
    \k_2^\top \K_3^{-1}\mathbf{c} = \frac{1}{k_{11}^2-k_{12}^2}( (k_{11}k_{13} - k_{12}k_{23})c_1 + (k_{11}k_{23} - k_{12}k_{13})c_2 )
\end{equation*}
which is 
\begin{equation*}
\frac{1}{k_{11}^2-k_{12}^2}( (k_{11} -k_{21})k_{13}c_1 + (k_{11} - k_{21})k_{23}c_2 + k_{12}(k_{23}-k_{13})(c_1-c_2) )\,,
\end{equation*}
or, equivalently,
\begin{equation}\label{equ:allstuff}
    \k_2^\top \K_3^{-1}\mathbf{c} = \frac{k_{13}c_1 + k_{23}c_2}{k_{11}+k_{12}}  +   \frac{k_{12}(k_{23}-k_{13})(c_1-c_2)}{k_{11}^2-k_{12}^2} \,,
\end{equation}
Note that the denominator $k_{11}+k_{12}$ is (assuming $k$ is monotonically decreasing with distance) nothing but a measure of sampling density, meaning that we write the left term in the sum as
\begin{equation}
    \k_2^\top \K_3^{-1}\mathbf{c} = \k_2^\top\D^{-1}\mathbf{c}  +   \frac{k_{12}(k_{23}-k_{13})(c_1-c_2)}{k_{11}^2-k_{12}^2} \,,
\end{equation}
where $\D$ is our lumped covariance matrix. Therefore,
\begin{equation}
    \left|\k_2^\top \K_3^{-1}\mathbf{c} -   \k_2^\top \D^{-1}\mathbf{c}\right|\leq \frac{k_{12}}{k_{11}^2-k_{12}^2}\left|k_{23}-k_{13}\right|\left|c_1-c_2\right|
\end{equation}
Thus, while our lumping introduces error, it is bounded and has the correct asymptotic behaviour (see \reffig{lumped-covariance-didactic} when the training samples are far from one another ($k_{12}\rightarrow0$), when the test samples are far enough from the training samples ($k_{23}-k_{13}\rightarrow0$), and when the training samples are close enough that the training features are close ($c_1-c_2\rightarrow 0$). Assuming independent samples \emph{without} accounting for sampling density would have meant approximating $k_{11}+k_{12}$ in \refequ{allstuff} by a constant factor, instead of recognizing it as a measure of the relative positions of $x_1$ and $x_2$. This would have introduced further error and abandoned the asymptotic convergence.

\end{document}